\documentclass[aps,prd,twocolumn,lengthcheck,tightenlines,superscriptaddress]{revtex4}

\usepackage{graphicx}
\usepackage{amsmath}
\usepackage{bm}
\usepackage{amsfonts}
\usepackage[pdftex,hidelinks]{hyperref}

\usepackage[utf8]{inputenc}
\usepackage[caption = false]{subfig}
\usepackage{placeins}
\usepackage{xcolor}
\usepackage{mathtools}
\usepackage{physics}
\usepackage{amssymb}
\usepackage{slashed}
\usepackage{makecell}

\newcommand{\jpsi}{{J/\psi}}
\newcommand{\chic}[1]{{\chi_{c#1}}}

\newcommand{\CS}[3]{{{}^{#1}#2_{#3}^{[1]}}}
\newcommand{\CO}[3]{{{}^{#1}#2_{#3}^{[8]}}}

\newcommand{\vt}[1]{\boldsymbol{#1}_\perp}
\newcommand{\vtp}[1]{\boldsymbol{#1}'_\perp}
\newcommand{\as}{\alpha_{\mathrm{s}}}
\begin{document}

\date{\today}

\author{Zhi-guo He}
\email{zhiguo.he@buct.edu.cn}
\affiliation{College of Mathematics and Physics, Beijing University of Chemical Technology, Beijing 100029, China}
\author{Hao-yu Liu}
\email{liuhy1@buct.edu.cn}
\affiliation{College of Mathematics and Physics, Beijing University of Chemical Technology, Beijing 100029, China}

\title{Forward $\chic{}$ production in $pp$ and $pA$ collisions within the CGC+NRQCD framework
}

\begin{abstract}
We study the forward $\chi_{cJ}$ production in the low $p_{\perp}$ region in proton-proton $(pp)$ and proton-nucleus $(pA)$ collisions at the LHC within the framework of non-relativistic QCD (NRQCD) factorization and the Color Glass Condensate (CGC) effective field theory. Incorporating the small-$x$ gluonic evolution via the running-coupling Balitsky-Kovchegov equation with various initial conditions, we systematically analyze the cross section ratio $R_{\chi_{c}}=\sigma_{\chi_{c2}}/\sigma_{\chi_{c1}}$, and find that the theoretical predictions weakly depend on these inputs. Our predictions align well with the LHCb measurements for $p\mathrm{Pb}$ collisions, however, overshoot those for $pp$ collisions. We further investigate the feed down fractions $F_{\chi_{c1,2}\to J/\psi}=\sigma_{\chi_{c1,2}\to J/\psi}/\sigma_{J/\psi}$. We notice that with the experimental measurement of prompt $J/\psi$ cross section as inputs in the denominator the CGC+NRQCD predictions broadly match the data when choosing the standard McLerran-Venugopalan initial conditions. In contrast, implementing purely theoretical CGC+NRQCD predictions for both $J/\psi$ and $\chi_{c}$ production yield systematic discrepancies across all tested parameters. This highlights a persistent challenge in achieving a simultaneous, unified description of $\chi_{c}$ and prompt $J/\psi$ production within the current framework.
\end{abstract}

\maketitle

\section{Introduction}

Heavy-quarkonium production provides a unique testing ground for studying the interplay of 
perturbative and nonperturbative QCD~\cite{Brambilla:2010cs}. The large heavy-quark mass 
introduces a hard scale that allows the short-distance production of a heavy-quark pair to be 
separated from its subsequent formation into a quarkonium state. Moreover, heavy quarkonium 
production can also probe initial-state effects, including nonlinear QCD dynamics in the small-Bjorken-$x$ regime~\cite{Kang:2013hta}.

The effective field theory of nonrelativistic QCD (NRQCD) offers a systematic framework for 
organizing this separation~\cite{Bodwin:1994jh}, in which the production cross section is 
expressed as a sum of perturbatively calculable short-distance coefficients multiplied by long-distance matrix elements (LDMEs) encoding the nonperturbative transition into a physical 
quarkonium state. Unlike the traditional color-singlet (CS) model, in NRQCD factorization the 
intermediate heavy quark pairs can be in both CS and color-octet (CO) states, whose 
contributions are weighted by powers of the relative velocity, $v$, of the heavy quark 
pair~\cite{Caswell:1985ui,Lepage:1992tx}. Combined with the collinear parton model (CPM), NRQCD 
factorization has achieved great success in describing quarkonium transverse momentum 
($p_\perp$) spectra in high-$p_\perp$ 
regions~\cite{Gong:2008ft,Ma:2010yw,Ma:2010jj,Butenschoen:2010rq,Butenschoen:2011yh}. 
However, the CPM+NRQCD calculations face notable limitations at low and moderate $p_\perp$. At 
high collision energies, the forward production of heavy quarkonium in such a lower-$p_\perp$ 
region opens direct access to extremely small values of $x$, where the growth of gluon 
densities towards small $x$ can enhance coherent multiple scattering. At sufficiently high 
densities, nonlinear gluon recombination tames this growth, leading to gluon 
saturation~\cite{Gribov:1983ivg,Mueller:1985wy,Mueller:1989st,McLerran:1993ni,McLerran:1993ka,Gelis:2010nm}. 

The Color Glass Condensate (CGC) effective field theory is a highly appropriate framework to 
describe these dynamics by separating the hadronic degrees of freedom into static, large-$x$ 
color sources and dynamical, small-$x$ gauge 
fields~\cite{Gelis:2010nm,Iancu:2003xm,Kovchegov:2012mbw,Albacete:2014fwa,Blaizot:2016qgz,Morreale:2021pnn}. Within this formulation, physical observables are obtained by averaging over a 
gauge-invariant source distribution whose evolution towards smaller values of $x$ is governed 
by the non-linear Balitsky--JIMWLK hierarchy~\cite{Balitsky:1995ub,Kovchegov:1999yj,Jalilian-Marian:1997qno,Jalilian-Marian:1997ubg,Iancu:2000hn,Ferreiro:2001qy}. 
This evolution dynamically generates a semi-hard saturation scale, $Q_s(x)$, which grows 
rapidly as $x$ decreases or as the nuclear size expands. Crucially, when this scale becomes 
significantly larger than the intrinsic QCD scale ($Q_s^2 \gg \Lambda_{\mathrm{QCD}}^2$), the 
asymptotic freedom of QCD ensures that the running coupling becomes weak [$\as(Q_s^2) \ll 1$], 
thereby rendering the otherwise non-perturbative low-$p_\perp$ domain accessible to systematic 
perturbative calculations.

To connect this CGC description of small-$x$ QCD dynamics with quarkonium observables, the 
short-distance production of a heavy quark pair must be supplemented by a treatment of its 
nonperturbative hadronization. Prior to full hadroproduction studies, the CGC framework was 
successfully deployed to investigate open heavy quark pair 
production~\cite{Blaizot:2004wv,Fujii:2006ab}. When subsequently combined with the color 
evaporation model~\cite{Fritzsch:1977ay,Halzen:1977rs}, comprehensive evaluations of 
$J/\psi$ and $\Upsilon$ yields were achieved~\cite{Fujii:2013gxa,Ducloue:2015gfa}. 
Meanwhile, within the CGC+NRQCD approach, combining pair production amplitudes with NRQCD 
projections led to analytic short distance coefficients (SDCs) for both $S$- and $P$-wave 
channels~\cite{Kang:2013hta}. Subsequent phenomenological applications of the CGC+NRQCD 
framework have primarily addressed the $S$-wave vector state $J/\psi$, including its production 
cross sections~\cite{Ma:2014mri,Ma:2015sia,Gimeno-Estivill:2024gbu} and 
polarization~\cite{Ma:2018bax}. In contrast, the $P$-wave charmonium states $\chi_{cJ}$ have 
received far less attention within this framework.

Phenomenologically, a dedicated investigation of forward $\chic{J}$ production is of 
significant interest. On the one hand, their hadroproduction is an essential ingredient of the 
feed down contribution to prompt $\jpsi$ yields~\cite{Yan-QingMa:2010}. On the other hand, 
their own production offers a direct test of $P$-wave heavy quarkonium production mechanism
~\cite{LHCb:2012chicFeeddown}, probing the interplay between the CS $\CS3PJ$ and CO $\CO3S1$ 
channels, which both contribute at leading order (LO) in the NRQCD velocity 
expansion~\cite{Bodwin:1994jh}. Furthermore, from the CGC perspective, $\chic{J}$ production 
provides a complementary way to study the multipoint color correlations that arise in the 
target color field. These correlations enter the CS channel through a quadrupole Wilson-line 
correlator~\cite{Kang:2013hta}, so their effect on the total cross section depends in part on 
the relative weight of this channel. The comparatively small CS fractions found in previous 
$\jpsi$ calculations~\cite{Ma:2014mri,Gimeno-Estivill:2024gbu} motivate examining the CS 
contribution and the dependence on multipoint-correlator prescriptions in $\chic{J}$ production.

In this work, we investigate forward $\chic{J}$ production at low transverse momentum in $pp$ 
and $pA$ collisions within the CGC+NRQCD framework~\cite{Kang:2013hta}. We first examine
different treatments for the multipoint correlators. Employing running-coupling Balitsky--Kovchegov (rcBK) evolution with various initial conditions, we then test predictions for the 
cross-section ratio $R_{\chi_{c}}=\sigma_{\chi_{c2}}/\sigma_{\chi_{c1}}$ against forward LHCb 
measurements~\cite{LHCb:2021zwp,LHCb:2013chicRatio}. We also analyze the feed down fractions 
$F_{\chic{1,2}\to \jpsi}=\sigma_{\chic{1,2}\to \jpsi}/\sigma_{\jpsi}$ in comparison with LHCb 
data~\cite{LHCb:2023chicFeeddown,LHCb:2012chicFeeddown} using two complementary schemes: a data 
driven approach normalizing our $\chic{}$ predictions by the measured prompt $\jpsi$ cross 
section, and a pure theoretical approach in which both the numerator and the 
denominator of prompt $\jpsi$ production are calculated within the same CGC+NRQCD framework, 
using six representative $\jpsi$ LDME sets. These joint evaluations serve as a highly 
sensitive probe of the consistency of the CGC+NRQCD framework across different charmonium 
states.

The remainder of this paper is organized as follows. In Sec.~\ref{sec:theory}, we describe the 
CGC+NRQCD factorization framework and discuss various approaches to handle the target 
Wilson-line correlators. In Sec.~\ref{sec:numeric}, we present the numerical results and 
compare them with the experimental measurements of the ratio $R_{\chi_c}$ 
and the feed  down fraction of $\chic{}$ to prompt $J/\psi$ hadroproduction. Finally, Sec.~\ref{sec:conclusions} is devoted to a summary of our conclusions.


\section{Theoretical Framework}\label{sec:theory}
 
\begin{figure*}[htbp]
    \centering
    \subfloat[]{
        \includegraphics{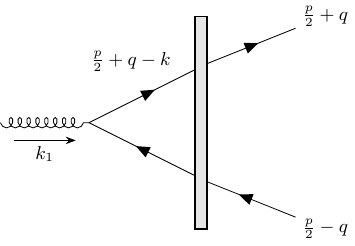}
        \label{fig:feynman_left}
    }
    \hspace{2cm}
    \subfloat[]{
        \includegraphics{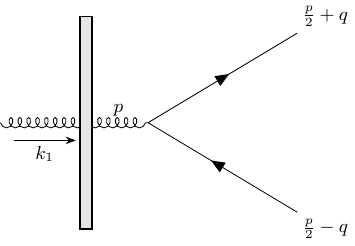}
        \label{fig:feynman_right}
    }
    \caption{Feynman diagrams for heavy quark--antiquark pair production where the projectile gluon splits before (a) or after (b) interacting with the target color field. The shaded bands represent the target shockwave.}
    \label{Fig: quarkonium production}
\end{figure*}

According to NRQCD factorization and framework of CGC, the differential cross section for the 
forward production of a quarkonium state $H$ in proton-proton ($pp$) or proton-nucleus ($pA$) 
collisions can be expressed as~\cite{Bodwin:1994jh,Kang:2013hta}
\begin{equation}
    \frac{\dd\sigma_{H}}{\dd^2\vt{p} \dd y} = \sum_\kappa \frac{\dd\hat{\sigma}^{\kappa}}{\dd^2\vt{p} \dd y} \langle \mathcal{O}_{\kappa}^{H}\rangle \,,
    \label{eq:cross_section}
\end{equation}
where $\frac{\dd\hat{\sigma}^{\kappa}}{\dd^2\vt{p} \dd y}$ is the SDC computed within CGC in 
the dilute–dense regime~\cite{Kang:2013hta}, and $\langle \mathcal{O}_{\kappa}^{H}\rangle$ 
is the NRQCD LDME with $\kappa={}^{2S+1}L_J^{[c]}$ denoting the intermediate $Q\bar{Q}$ pair in 
state $\kappa$, carrying the quantum numbers of spin $S$, orbital angular momentum $L$, and 
total angular momentum $J$. The color configuration is represented by $c=1$ for CS and $c=8$ 
for CO. Under the NRQCD velocity scaling rules, the dominant intermediate states for $\jpsi$ 
include $\CS3S1$, $\CO1S0$, $\CO3S1$, and $\CO3PJ$, while for $\chic{J}$ the leading 
contributions arise from the $\CO3S1$ and $\CS3PJ$ states. In this work, we adopt collinear 
factorization for the dilute projectile proton and the CGC for the dense target. At QCD LO, 
only the gluon channel contributes via the partonic process $g(k_1)+A\to Q\bar Q(p)+X$.   As 
illustrated in Fig.~\ref{Fig: quarkonium production}, the projectile gluon can split into the 
heavy quark pair either before or after crossing the target shockwave. Here, $p$ is the total 
momentum of the pair, and $q$ denotes the relative momentum between them.

We work in the nucleon--nucleon center-of-mass frame, where the projectile proton moves along 
$+z$ with momentum $p_p$ and the target along $-z$ with momentum $p_A$ (defined per nucleon of a nuclear target). In the light-cone coordinates, $v^\pm=(v^0\pm v^3)/\sqrt{2}$, the four vectors can be written as $(v^+,v^-,\vt{v})$, and we choose the light-cone gauge $n\cdot A=A^+=0$ with $n^\mu=(0,1,\vt{0})$. In the high-energy limit, the longitudinal components satisfy $p_p^+=p_A^-=\sqrt{s/2}$, where $s=(p_p+p_A)^2$ is the squared center-of-mass energy per nucleon pair. In line with the collinear treatment of the projectile, the incoming gluon carries momentum $k_1=(x_p p_p^+,0,\vt{0})$. In the non-relativistic limit, the quarkonium mass is twice the heavy quark mass $m$. The projectile momentum fraction $x_p$ and the typical target momentum fraction $x_A$ are then given by $x_{p,A}=\sqrt{4m^2+\vt{p}^2}\,e^{\pm y}/\sqrt{s}$, where $y=\frac{1}{2}\ln(p^+/p^-)$ is the rapidity of the heavy quark pair and the upper (lower) sign corresponds to the projectile (target).

In these conventions, the heavy quark pair production amplitude can be read from the Feynman diagrams in Fig.~\ref{Fig: quarkonium production} directly
\begin{align}\label{eq:QQamp}
\begin{split}
\mathcal{M}(p,q) &= \frac{g_{\mathrm{s}}}{(2\pi)^2} \int \dd^2\vt{k} \, \dd^2\vt{x} \, \dd^2\vt{y} e^{i\vt{k}\cdot\vt{x}} e^{i (\vt{p}-\vt{k})\cdot\vt{y}} \\
&\; \times \bar{u}\left(\frac{p}{2}+q\right) \bigg[ T_{q\bar{q}}\left(p,q,\vt{k}\right) U(\vt{x}) t^a U^\dagger(\vt{y}) \\
&\; + T_g(p) t^b W^{ba}(\vt{x}) \bigg] v\left(\frac{p}{2}-q\right).
\end{split}
\end{align}
In diagram~(a), $\vt{x}$ and $\vt{y}$ denote the quark and antiquark transverse coordinates at 
the target, and $\vt{k}$ is the transverse component of the momentum $k$ transferred from the 
target to the quark. In diagram~(b), $\vt{x}$ instead corresponds to the transverse coordinate 
of the gluon crossing the target. The hard scattering structures $T_{q\bar{q}}$ and $T_g$ are 
given by 
\begin{widetext}
\begin{align}
T_{q\bar{q}}\left(p,q,\vt{k}\right) &= \frac{\slashed{n} \left( \frac{\slashed{p}}{2} + \slashed{q} - \slashed{k} + m\right) \slashed{\epsilon}(k_1) \left( \frac{\slashed{p}}{2} + \slashed{q} - \slashed{k} - \slashed{k}_1 + m\right) \slashed{n}} {2p^+\left[\left(\frac{\vt{p}}{2}+\vt{q}-\vt{k}\right)^2+m^2\right]}, \\
T_g(p) &= \frac{ \epsilon_\mu(k_1) \Lambda^{\mu\nu\rho}(k_1, -p, p-k_1) n_\rho d_{\nu\sigma}(p) \gamma^\sigma }{p^2} \;.
\end{align}
Here, $\epsilon^\mu(k_1)$ is the incoming-gluon polarization vector, 
$\Lambda^{\mu\nu\rho}$ represents the three-gluon vertex, and $d_{\nu\sigma}(p)=-g_{\nu\sigma}+(p_\nu n_\sigma+n_\nu p_\sigma)/(p\cdot n)$ is the numerator of gluon propagator  in this gauge.

The Wilson lines $U$ and $W$ in Eq.~\eqref{eq:QQamp} encode the multiple eikonal scattering of quarks and gluons in the target color field, respectively. The fundamental Wilson line is given by
\begin{equation}\label{eq:wilsonlinef}
U\left(\vt{r}\right) = \mathcal{P} \exp \left\{i g_{\mathrm{s}} \int_{-\infty}^{+\infty} \dd r^{+} t^c A_c^{-}\left(r^{+}, \vt{r}\right)\right\} \;.
\end{equation}
Here, $t^c$ are the $SU(3)$ generators in the fundamental representation, $\mathcal{P}$ denotes path ordering along $r^+$, and $A_c^{-}(r^{+},\vt{r})$ is the classical color field of the target. The adjoint Wilson line $W^{ba}(\vt{x})$ is obtained by replacing the fundamental generators in Eq.~\eqref{eq:wilsonlinef} with those in the adjoint representation.

With the help of the spin projection method~\cite{Kang:2013hta}, the amplitude can be projected onto the $S$-wave and $P$-wave Fock states straightforwardly. After squaring the projected amplitude, averaging over the target color sources, and convolving the result with the collinear gluon parton distribution function (PDF) $f_{g/p}(x_p,\mu^2)$ of the projectile proton, we obtain the short-distance cross section
\begin{align}\label{eq:cross_section_partonic}
\begin{split}  
\frac{\dd\hat{\sigma}^\kappa}{\dd^2\vt{p} \dd y} &= \frac{\as}{4(2\pi)^5(N_c^2-1)} x_p f_{g/p}(x_p,\mu^2)  \int \dd^2\vt{k} \dd^2\vtp{k} \int\dd^2\vt{x} \, \dd^2\vt{y} \, \dd^2\vtp{x} \, \dd^2\vtp{y}   \, \\\quad \times&e^{i\big[ \vt{k}\cdot\vt{x} - \vtp{k}\cdot\vtp{x} + (\vt{p}-\vt{k})\cdot\vt{y} - (\vt{p}-\vtp{k})\cdot\vtp{y} \big]} \mathcal{C}^\kappa(\vt{x}, \vtp{x}, \vtp{y}, \vt{y}) W^\kappa(\vt{k},\vtp{k}) \;.
\end{split}
\end{align}
\end{widetext}
Here, primed transverse momenta and coordinates refer to the complex-conjugate amplitude, 
$W^\kappa$ is the hard factor in momentum space for the Fock state $\kappa$, and 
$\mathcal{C}^\kappa$ encodes the target correlations associated with its color configuration. 
Our results are in agreement with the expressions derived for the collinear proton in the
literature~\cite{Kang:2013hta}. 

For the CO contribution, the target color structure can be further factorized into a product of 
two fundamental dipole correlators in the large $N_c$ and large nucleus 
limit~\cite{Kang:2013hta}:
\begin{equation}\label{eq:color_octet_factorized}
\mathcal{C}^{[8]}(\vt{x},\vtp{x},\vtp{y},\vt{y})\simeq D(\vt{x}-\vtp{x})D(\vt{y}-\vtp{y}),
\end{equation}
where the fundamental dipole correlator, if the impact parameter dependence is neglected, is 
defined as
\begin{equation}
D\left(\vt{r}\right) = \frac{1}{N_c} \langle\operatorname{Tr}\left[U\left(\vt{r}\right) U^{\dagger}\left(\vt{0}\right)\right]\rangle_{X_f} \;,
\end{equation}
with $r_\perp=|\vt{r}|$. The brackets $\langle\cdots\rangle_{X_f}$ denote the CGC ensemble 
average over target color source configurations at the rapidity factorization scale $X_f$. The 
evolution of $D(r_\perp)$ toward smaller value of $X_f$ is described by the Balitsky--Kovchegov (BK) equation~\cite{Balitsky:1995ub,Kovchegov:1999yj}, obtained in the large $N_c$ mean-field 
limit.

The factorized color structure in Eq.~\eqref{eq:color_octet_factorized}, together with the 
assumed independence of impact parameter, allows us to perform the integration over the 
transverse coordinates in Eq.~\eqref{eq:cross_section_partonic} analytically. This enables us 
to factor out the effective transverse area of the target, $S_\perp = \int \dd^2 \vt{b}$, 
yielding:
\begin{align}\label{eq:fourier}
\begin{split}
&\int \dd^2\vt{x} \, \dd^2\vt{y} \, \dd^2\vtp{x} \, \dd^2\vtp{y} \, \\&\quad \times e^{i\big[ \vt{k}\cdot\vt{x} - \vtp{k}\cdot\vtp{x} + (\vt{p}-\vt{k})\cdot\vt{y} - (\vt{p}-\vtp{k})\cdot\vtp{y} \big]} \\
&\quad \times D(\vt{x}-\vtp{x})D(\vt{y}-\vtp{y})  \\
&= (2\pi)^2 S_\perp \mathcal{N}(\vt{k}) \mathcal{N}(\vt{p}-\vt{k}) \delta^{(2)}(\vt{k}-\vtp{k}) \;,
\end{split}
\end{align}
where $\mathcal{N}(\vt{k})$ is the Fourier transform of the fundamental dipole correlator:
\begin{equation}\label{eq:dipolefourierq}
 \mathcal{N}(\vt{k}) = \int \dd^2 \vt{r} \, e^{-i\vt{k}\cdot \vt{r}} D(\vt{r}) \;.
\end{equation}

After integrating out $\vtp{k}$ via the delta function, the CO hard factor can be written in a simpler form as $W_8^\kappa(\vt{k})\equiv W^\kappa(\vt{k},\vt{k})$, and the corresponding CO short-distance cross section becomes
\begin{align}   
\frac{\dd\hat{\sigma}_{\mathrm{CO}}^\kappa}{\dd^2\vt{p} \dd y}  &= \frac{\as S_{\perp}}{4(2\pi)^3(N_c^2-1)} x_p f_{g/p}(x_p,\mu^2) \nonumber \\
&\times \int \dd^2\vt{k} \, \mathcal{N}(\vt{k}) \mathcal{N}(\vt{p}-\vt{k}) W_8^\kappa(\vt{k}) \;.
    \label{eq:cross_section_octet}
\end{align}
The CO cross section can therefore be evaluated efficiently, since $\mathcal{N}(\vt{k})$ follows directly from the Fourier transform of the BK-evolved dipole correlator.

\begin{widetext}
By contrast, taking the color trace for the CS contribution leads to a more intricate structure, whose finite $N_c$ form is given by~\cite{Kang:2013hta}:
\begin{equation}\label{eq:color_singlet_finite_nc}
\mathcal{C}^{[1]}(\vt{x},\vtp{x},\vtp{y},\vt{y})=Q(\vt{x},\vtp{x},\vtp{y},\vt{y})-S^{(2,2)}(\vt{x},\vt{y};\vtp{y},\vtp{x}),
\end{equation}
where $Q=\frac{1}{N_c}\left\langle\operatorname{Tr}\left[U(\vt{x})U^\dagger(\vtp{x})U(\vtp{y})U^\dagger(\vt{y})\right]\right\rangle_{X_f}$ and $S^{(2,2)}=\frac{1}{N_c^2}\left\langle\operatorname{Tr}\left[U(\vt{x})U^\dagger(\vt{y})\right]\operatorname{Tr}\left[U(\vtp{y})U^\dagger(\vtp{x})\right]\right\rangle_{X_f}$ denote the normalized single-trace quadrupole and double-trace correlators, respectively. In the large $N_c$ and large-nucleus limit, $S^{(2,2)}$ further factorizes into a product of two dipole correlators, simplifying Eq.~\eqref{eq:color_singlet_finite_nc} to
\begin{equation}\label{eq:color_singlet_factorized}
\mathcal{C}^{[1]}(\vt{x},\vtp{x},\vtp{y},\vt{y})\simeq Q(\vt{x},\vtp{x},\vtp{y},\vt{y})-D(\vt{r})D(\vtp{r}),
\end{equation}
where $\vt{r}=\vt{x}-\vt{y}$ and $\vtp{r}=\vtp{x}-\vtp{y}$ are the transverse separations between the heavy quark and antiquark in the amplitude and complex-conjugate amplitude, respectively.
\end{widetext}

Although the quadrupole appearing in Eq.~\eqref{eq:color_singlet_factorized} is governed by the 
Balitsky-JIMWLK hierarchy just as the dipole correlator, its numerical solution remains much 
more challenging than solving only the BK equation~\cite{Alvioli:2012Gaussian}, even within the 
large $N_c$ mean-field approximation. In practice, various approximate approaches have been 
developed to evaluate such CS multipoint correlators. Here we consider two approaches: a dipole 
based ansatz and the Gaussian approximation. The dipole based ansatz expresses the quadrupole 
$Q$ completely in terms of fundamental dipole correlators, which enables an analytic Fourier 
transformation of the CS target structure in Eq.~\eqref{eq:color_singlet_factorized} analogous 
to Eq.~\eqref{eq:fourier}, expressing the target dependence through convolutions of the 
momentum-space dipole profile $\mathcal{N}(\vt{k})$. Thus the CS short-distance cross section 
can be written as
\begin{align}\label{eq:dsktCS1}
\frac{\dd \hat{\sigma}_{\mathrm{CS}}^\kappa}{\dd^2\vt{p} \dd y}  &= \frac{\as S_\perp}{4(2\pi)^{5} (N_c^2-1)} \int \dd^2\vt{k} \, \dd^2\vtp{k} \nonumber\\
&\quad \times x_{p} f_{g/p}\left(x_{p}, \mu^{2}\right) \tilde{\mathcal{H}}^\kappa_1 \;.
\end{align}
The explicit expression of the hard kernel $\tilde{\mathcal{H}}^\kappa_1$ is presented in Appendix~\ref{app:quadrupole-approximation}.

\begin{table*}[!htbp]
\caption{The values of NRQCD LDMEs we used for $\jpsi$ production in units of $\mathrm{GeV}^3$. }
\label{tab:ldme}
\centering
\footnotesize
\renewcommand{\arraystretch}{1}
\setcellgapes{3pt}
\makegapedcells
\setlength{\tabcolsep}{5pt}
\begin{tabular}{|c|c|c|c|c|c|c|}
\hline
 & Set~1~\cite{He:2026hqss}
 & Set~2~\cite{Brambilla:2024iqg}
 & Set~3~\cite{Chao:2012iv}
 & Set~4~\cite{Bodwin:2014gia}
 & Set~5~\cite{Butenschoen:2011yh}
 & Set~6~\cite{Gong:2012ug} \\
\hline
$\left\langle \mathcal{O}^{\jpsi}(\CS3S1)\right\rangle$
 & $1.16/6$ & $1.16/6$ & $1.16/6$ & $1.16/6$ & $1.32/6$ & $1.16/6$ \\
\hline
$\left\langle \mathcal{O}^{\jpsi}(\CO1S0)\right\rangle$
 & $0.0014$ & $0.00068$ & $0.089$ & $0.099$ & $0.0304$ & $0.097$ \\
\hline
$\left\langle \mathcal{O}^{\jpsi}(\CO3P0)\right\rangle/m^2$
 & $0.0186$ & $0.01879$ & $0.0056$ & $0.00489$ & $-0.00404$ & $-0.0095$ \\
\hline
$\left\langle \mathcal{O}^{\jpsi}(\CO3S1)\right\rangle$
 & $0.0109$ & $0.0105$ & $0.0030$ & $0.0110$ & $0.00168$ & $-0.0046$ \\
\hline
\end{tabular}
\end{table*}

In the Gaussian approximation, the multipoint Wilson-line correlators are derived from a common 
Gaussian ensemble of target fields and are expressed in terms of the fundamental dipole 
correlator~\cite{Fujii:2006ab,Marquet:2010Gaussian,Iancu:2011Gaussian}. In contrast to the 
dipole based ansatz, the Gaussian expressions do not allow for a reduction to convolutions 
involving only momentum space dipole profiles. Substituting them into 
Eq.~\eqref{eq:cross_section_partonic} retains the target correlators in coordinate space, 
where the highly oscillatory Fourier kernels make a direct numerical integration 
impractical. To overcome this difficulty, we first perform the integration over the internal 
momenta and then rewrite the cross section as
\begin{align}\label{eq:cross_section_singlet_gauss}   
\frac{\dd\hat{\sigma}_{\mathrm{CS}}^\kappa}{\dd^2\vt{p} \dd y} &= \frac{\as S_{\perp}}{4(2\pi)^3(N_c^2-1)} x_p f_{g/p}(x_p,\mu^2) \nonumber\\
&\quad \times \int \dd^2\vt{\Delta} \, \dd^2\vt{r} \, \dd^2\vtp{r} \, e^{-i\vt{p} \cdot \vt{\Delta}} \nonumber\\
&\quad \times \mathcal{C}_{\mathrm{G}}^{[1]} \Gamma_1^\kappa(\vt{r},\vtp{r}) \;,
\end{align}
where $\vt{\Delta} = (\vtp{x}+\vtp{y}-\vt{x}-\vt{y})/2$. The color structure 
$\mathcal{C}_{\mathrm{G}}^{[1]}$ is evaluated by using Eq.~\eqref{eq:color_singlet_finite_nc} 
for the finite $N_c$ Gaussian prescription and Eq.~\eqref{eq:color_singlet_factorized} for the 
large $N_c$ prescription. The explicit expressions of the quadrupole term, $Q(\vt{x},\vtp{x},\vtp{y},\vt{y})$, and the double trace term, $S^{(2,2)}(\vt{x},\vt{y};\vtp{y},\vtp{x})$, are available in literature~\cite{Dominguez:2011wm,Cheung:2024qvw}. These expressions are evaluated with the rcBK-evolved dipole as input at each $X_f$. The coordinate space hard factor 
$\Gamma_{1}^\kappa(\vt{r},\vtp{r})$, with its implicit dependence on $\vt{p}$, is related to the Fourier transform of the hard part in momentum space as:
\begin{equation}\label{eq:Gamma1_def}
\begin{aligned}
\Gamma_{1}^{\kappa} &\equiv \frac{1}{(2 \pi)^{2}} \int \dd^2\vt{l} \, \dd^2\vtp{l} \\
&\quad \times e^{i\vt{l} \cdot \vt{r}} e^{-i\vtp{l} \cdot \vtp{r}} W^\kappa(\vt{l},\vtp{l}) \;,
\end{aligned}
\end{equation}
with the shifted momenta $\vt{l} = \vt{k} - \frac{\vt{p}}{2}$ and 
$\vtp{l} = \vtp{k} - \frac{\vt{p}}{2}$. Since the hard parts $W^\kappa$ are sums of products of 
functions depending separately on $\vt{l}$ and $\vtp{l}$, the two Fourier transforms can be 
performed independently. In our computation, we perform the momentum integrations in 
$\Gamma_1^\kappa$ analytically, thereby reducing the dimensionality and separating the two 
oscillatory Fourier integrals from the numerical evaluation, which significantly improves the 
numerical stability. These expressions are collected in Appendix~\ref{app:Gamma-singlet}.

\section{Numerical Analysis}\label{sec:numeric}
Before comparing with the experimental measurements, we first specify the required numerical 
inputs. For the collinear projectile proton, we use the central member of the CT18 NLO PDF 
set~\cite{Hou:CT18} and its associated running coupling $\as$ for the hard partonic process, 
both accessed through LHAPDF6~\cite{Buckley:LHAPDF6}. The target dipole correlator is 
obtained by numerically solving the rcBK 
equation~\cite{Kovchegov:2006wf,Kovchegov:2006vj,Balitsky:2006wa,Albacete:2010sy}. We assess 
the initial condition dependence using three parameter sets, labeled CGC sets~1, 2, and 3, for 
the generalized McLerran--Venugopalan (MV) form at the initial evolution scale $X_0=0.01$~\cite{McLerran:1993ni,McLerran:1993ka,McLerran:1994vd,McLerran:1997fk,Albacete:2010sy},
\begin{equation}\label{eq:MV}
D_{X_0}\left(r_\perp\right) = \exp \left[-\frac{\left(r_\perp^2 Q_{s 0}^2\right)^\gamma}{4} \ln \left(\frac{1}{r_\perp\Lambda_{\mathrm{QCD}} }+e\right)\right],
\end{equation}
where $Q_{s0}$ denotes the initial saturation scale, and $\Lambda_{\mathrm{QCD}}$ is fixed to 
$0.241\,\mathrm{GeV}$. CGC set~1 uses the standard MV initial condition, with $\gamma=1$ and 
$Q_{s0,p}^2=0.2\,\mathrm{GeV}^2$, while set~2 uses the g1.101 ($\gamma=1.101$) parametrization 
and set~3 uses g1.119 ($\gamma=1.119$). The latter two parametrizations are based on fits to 
HERA small-$x$ deep-inelastic-scattering data. For all three CGC sets, we use the initial 
saturation scales and running-coupling parameters given in Table~I of 
Ref.~\cite{Albacete:2012xq}. Across all three CGC sets, we use the homogeneous prescription 
$Q_{s0,A}^2=NQ_{s0,p}^2$, with $N=1$ for a proton and $N=2$ for lead ($A=208$), following 
Ref.~\cite{Ma:2015sia}. The same effective transverse areas are used: $S_{\perp,p}=\pi R_p^2$ 
with $R_p=0.48\,\mathrm{fm}$ for $pp$ collisions, and $S_{\perp,A}=\pi R_A^2$ with 
$R_A=\sqrt{A/2}\,R_p\simeq4.9\,\mathrm{fm}$ for $p\mathrm{Pb}$ 
collisions~\cite{Ma:2014mri,Ma:2015sia}. 

We set the central rapidity factorization scale $X_f$ to be $x_A$ and vary it over the interval 
$[x_A/2,\,2x_A]$ to estimate the uncertainties. The collinear factorization and renormalization 
scales are equated, $\mu_F=\mu_R=\mu=\xi\sqrt{p_\perp^2+4m^2}$ with charm quark mass 
$m=1.5\,\mathrm{GeV}$, and $\xi$ is varied from $1/2$ to 2. When calculating the ratios 
$R_{\chi_c}$ and $F_{\chic{1,2}\to \jpsi}$, we apply identical scale choices to both the 
numerator and denominator and determine separately the maximum upward and downward shifts 
induced by varying $X_f$ and $\xi$ independently, and combine these scale uncertainties in 
quadrature. For the NRQCD inputs, we follow the normalization convention of 
Ref.~\cite{Kang:2013hta}, in which at $v^2$ LO the P-wave CS LDMEs are related to the first 
derivative of wave functions at origin through
\begin{equation}\label{eq:ldme_chic_cs}
\langle \mathcal{O}^{\chic{J}}(\CS3PJ)\rangle = \frac{3(2J+1)}{4\pi} |R_P'(0)|^2 \;,
\end{equation}
and adopt the value, $|R_P'(0)|^2=0.075\,\mathrm{GeV}^{5}$, in potential model evaluation by 
using the Buchm\"uller--Tye potential~\cite{Eichten:1995PRD}. At $v^2$ LO, the relations 
among the CO LDMEs are
\begin{equation}\label{eq:ldme_chic_co}
\langle\mathcal{O}^{\chic{J}}(\CO3S1)\rangle=(2J+1)\langle\mathcal{O}^{\chic{0}}(\CO3S1)\rangle \;.
\end{equation}
We take the value $\langle\mathcal{O}^{\chic{0}}(\CO3S1)\rangle\simeq2.2\times10^{-3}\,\mathrm{GeV}^3$ 
determined from the QCD NLO fit to CDF data~\cite{Yan-QingMa:2010} using the same CS ones. For 
the theoretical prediction of direct $\jpsi$ production, we consider the six representative 
$\jpsi$ LDME sets listed in Table~\ref{tab:ldme}. In the end, the $\chi_c$ feed down branching 
function $\mathcal{B}(\chic1\to \jpsi\gamma)=(33.9\pm1.2)\%$ and $\mathcal{B}(\chic2\to \jpsi\gamma)=(19.0\pm0.8)\%$, can be found in the Particle Data Group~\cite{ParticleDataGroup:2026aaa}, and we neglect the $\chic{0}$ feed down contribution for its tiny branching ratio to $J/\psi$.

\begin{figure}[!htbp]
    \centering
    \subfloat[]{
        \includegraphics[width=0.45\columnwidth]{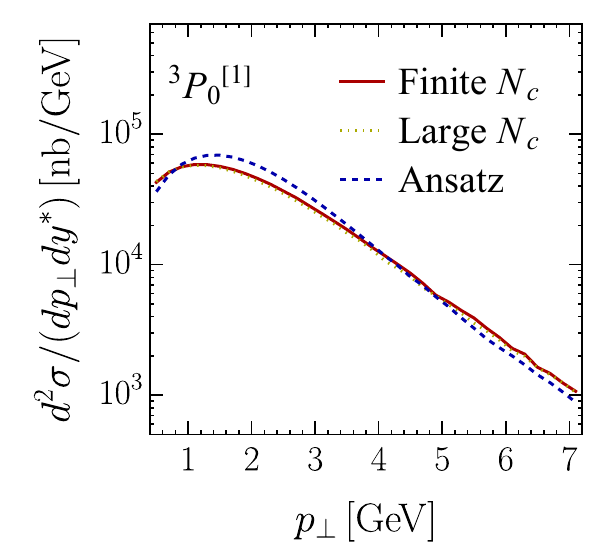}
        \label{fig:group2_a}
    }
    \hfill
    \subfloat[]{
        \includegraphics[width=0.45\columnwidth]{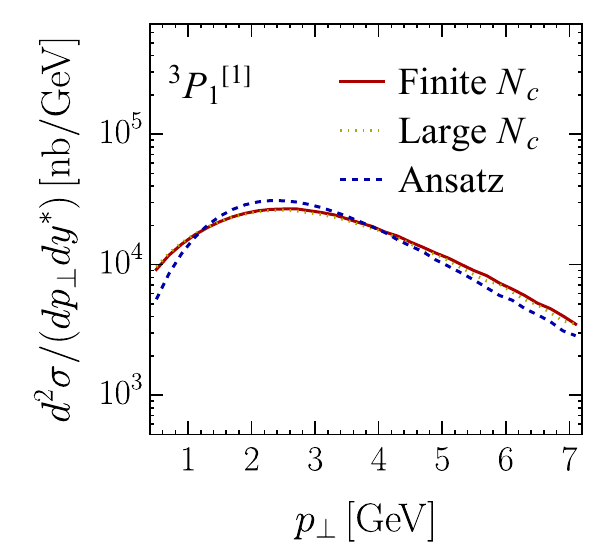}
        \label{fig:group2_b}
    }
    \\ 
    \vspace{0.2cm}
    \subfloat[]{
        \includegraphics[width=0.45\columnwidth]{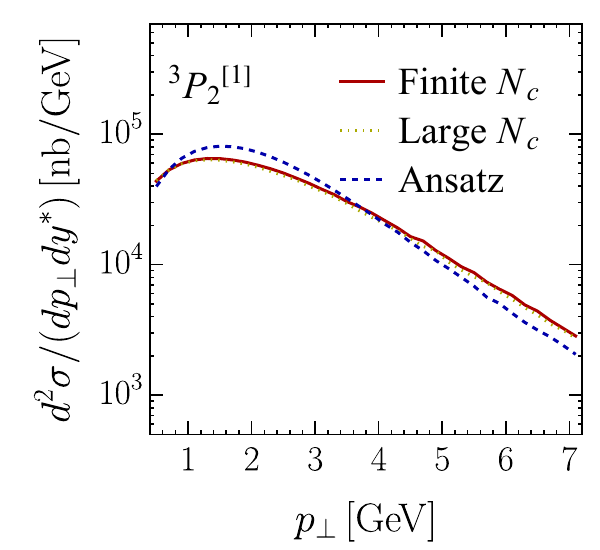}
        \label{fig:group2_c}
    }
    \hfill
    \subfloat[]{
        \includegraphics[width=0.45\columnwidth]{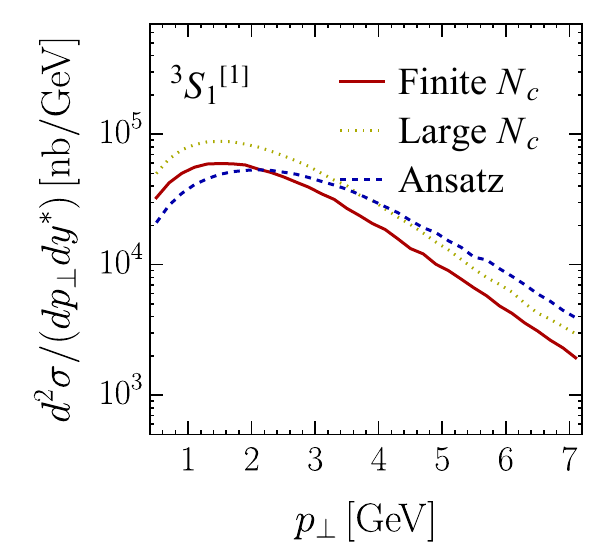}
        \label{fig:group2_d}
    }
    \caption{Transverse momentum spectra of the CS contributions for CGC set~1 with the finite $N_c$ Gaussian (solid red), large $N_c$ Gaussian (dotted yellow), and dipole based (dashed blue) prescriptions in $p\mathrm{Pb}$ collisions at $\sqrt{s_{NN}}=8.16\,\mathrm{TeV}$ in rapidity range $1.5<y<4.0$. Panels (a)--(d) show the contributions from the $\CS3P0$, $\CS3P1$, $\CS3P2$, and $\CS3S1$ channels, respectively.}
    \label{fig:group2_qqpair}
\end{figure}

To guide the subsequent phenomenology, we first investigate the properties of the three 
different prescriptions for the multipoint correlators addressed in Sec.~\ref{sec:theory}. 
Figure~\ref{fig:group2_qqpair} shows the CS contributions to the $\CS3PJ$ channels in
$\chic{J}$ production and the $\CS3S1$ channel for $\jpsi$ production. The two Gaussian 
prescriptions yield similar spectral shapes, but their magnitudes differ: the large $N_c$ 
approximation enhances the $\CS3S1$ yield by $35$--$55\%$ relative to the finite $N_c$ results 
over the displayed range; however, it can only lead to less than $10\%$ difference in the 
$\CS3PJ$ channel. The dipole based ansatz of Ref.~\cite{Ma:2014mri}, on the other hand, 
produces different spectral shapes, particularly at low and high $p_\perp$, with deviations 
from the finite $N_c$ Gaussian results around $40\%$ in the $\CS3PJ$ channels and $100\%$ in 
the $\CS3S1$ channel. Although the dipole based ansatz simplifies the momentum space 
calculation, the Gaussian approximation provides a common construction of the relevant 
multipoint correlators~\cite{Iancu:2011Gaussian}. The finite $N_c$ quadrupole expressions agree 
well with numerical JIMWLK results for the tested configurations~\cite{Dumitru:2011vk}. 
Retaining the finite $N_c$ color structure also leaves the integration dimensionality unchanged 
relative to the large $N_c$ limit. We therefore adopt the finite $N_c$ Gaussian prescription 
for the CS multipoint correlators in the following analysis.

\begin{figure}[!htbp]
    \centering
        \includegraphics[width=0.9\columnwidth]{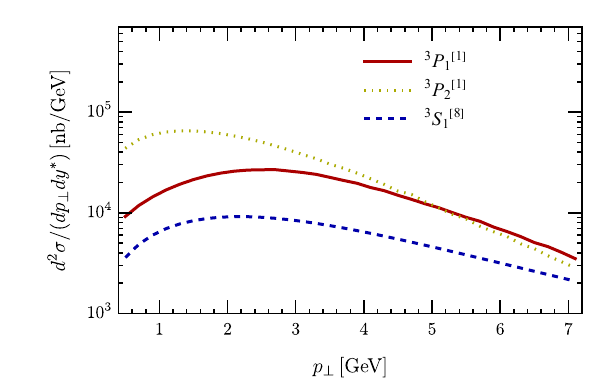}
    \caption{CS and CO contributions to $\chic1$ and $\chic2$ production, with the same CGC input and kinematics as in Fig.~\ref{fig:group2_qqpair}. The CS contributions use the finite $N_c$ Gaussian prescription. The $\CO3S1$ curve corresponds to $\chic1$ production, while the $\chic2$ contribution can be obtained by a normalization factor of $5/3$ from heavy-quark spin symmetry.}
    \label{fig:chicch}
\end{figure}

In Fig.~\ref{fig:chicch}, the CS and CO contributions to $\chic{1,2}$ production are presented 
individually. We find that with our numerical inputs the CS channel can make up approximately 
$45$--$90\%$ of $\chic2$ production and more than $60\%$ of $\chic1$ production over the 
displayed $p_\perp$ range. These sizable fractions stand in sharp contrast to the approximately 
$15\%$ CS contribution to the $p_\perp$-integrated $\jpsi$ cross section reported in 
Ref.~\cite{Gimeno-Estivill:2024gbu}, highlighting $\chic{1,2}$ production as a more sensitive 
probe of the multipoint color correlations entering the CS channel.

\begin{figure*}[!htbp]
    \centering
    \subfloat[]{
        \includegraphics[width=0.48\textwidth]{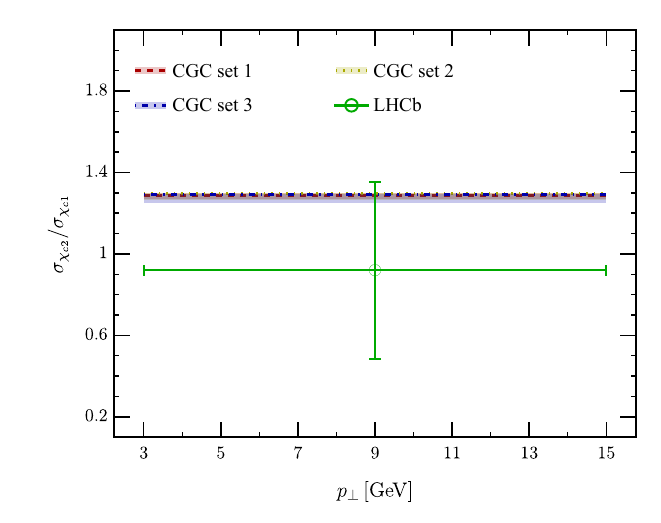}
        \label{fig:chic21_a}
    }
    \hfill
    \subfloat[]{
        \includegraphics[width=0.48\textwidth]{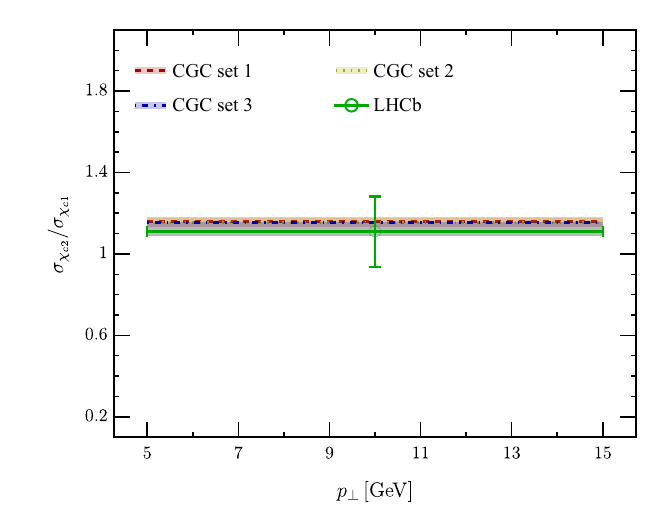}
        \label{fig:chic21_b}
    }
    \caption{Comparison of CGC+NRQCD prediction for $R_{\chi_c}$ ratios with the LHCb measurements using converted photons (a) and calorimetric photons(b) in forward $p\mathrm{Pb}$ collisions at $\sqrt{s_{NN}}=8.16\,\mathrm{TeV}$ within $1.5<y<4.0$~\cite{LHCb:2021zwp}. }
    \label{fig:chic21}
\end{figure*}

\begin{figure}[!htbp]
    \centering
    \includegraphics[width=0.90\columnwidth]{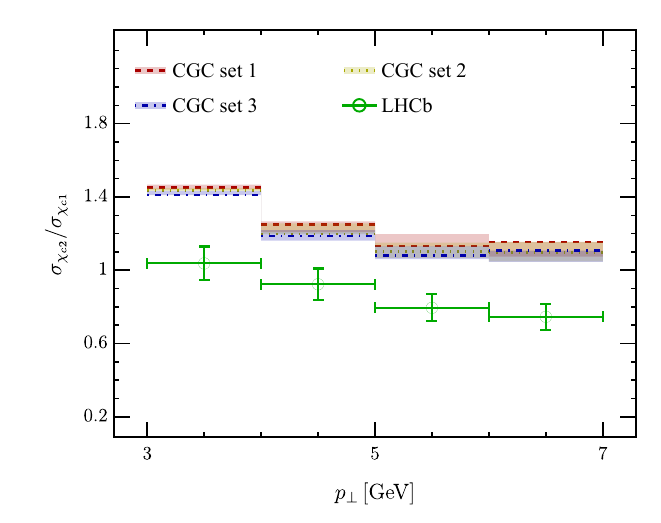}
    \caption{Comparison of CGC+NRQCD predictions of $R_{\chi_c}$ ratio with the LHCb measurements in forward $pp$ collisions at $\sqrt{s}=7\,\mathrm{TeV}$ within $2.0<y<4.5$~\cite{LHCb:2013chicRatio}.}
    \label{fig:chic21_pp}
\end{figure}

We next compare the ratio $R_{\chi_c}$ with forward LHCb measurements, beginning with 
$p\mathrm{Pb}$ collisions at $\sqrt{s_{NN}}=8.16\,\mathrm{TeV}$ within the rapidity range 
$1.5<y<4.0$~\cite{LHCb:2021zwp}, as shown in Fig.~\ref{fig:chic21}. To match the experimental 
kinematic windows, the cross sections are obtained by integrating over $3<p_\perp<15\,\mathrm{GeV}$ for the converted photon sample and $5<p_\perp<15\,\mathrm{GeV}$ for the 
calorimetric-photon sample. We notice that the ratio depends only weakly on the rcBK initial 
condition, resulting from substantial cancellation of the common dipole dependence. Partial 
cancellation of the scale dependence also reduces the uncertainty estimated through correlated 
variations of $\mu$ and $X_f$. Our predictions are consistent with the measurements within 
experimental uncertainties. In Fig.~\ref{fig:chic21_pp}, we display the corresponding 
comparison with the LHCb converted photon measurement in $pp$ collisions at $\sqrt{s}=7\,\mathrm{TeV}$ and $2.0<y<4.5$~\cite{LHCb:2013chicRatio}. Since the measurements are binned in 
the daughter $\jpsi$ transverse momentum, in our predictions, we approximately take 
$p_{\perp}^{\jpsi}\simeq p_{\perp}^{\chic{}}$. In contrast to $p\mathrm{Pb}$ collisions, 
predictions from all three CGC sets systematically overshoot the measured ratio over the 
displayed range, while yielding similar transverse momentum shapes. In CPM+NRQCD, NLO 
calculations lead to a substantially smaller value of $R_{\chi_c}$ than at LO~\cite{Yan-QingMa:2010,Shao:2014chicPolarizations}. These results strongly motivate a dedicated study of 
higher order perturbative corrections within CGC+NRQCD.

\begin{figure*}[!htbp]
    \centering
    \subfloat[]{
        \includegraphics[width=0.48\textwidth]{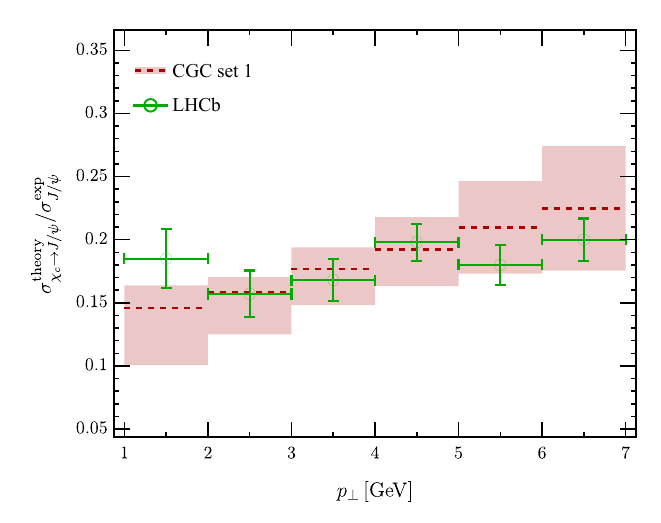}
        \label{fig:chicbr_data_pa}
    }
    \hfill
    \subfloat[]{
        \includegraphics[width=0.48\textwidth]{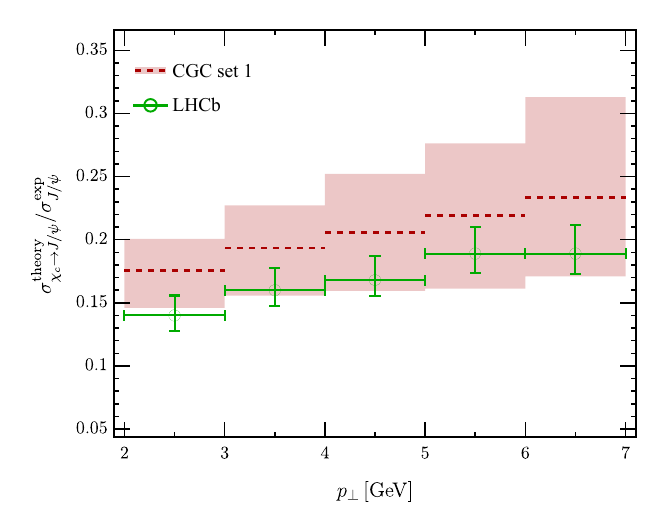}
        \label{fig:chicbr_data_pp}
    }
    \caption{Comparison of CGC+NRQCD predictions using CGC set~1 for the data driven feed down fractions $F_{\chic{1,2}\to \jpsi}$ with the LHCb measurements in (a) forward $p\mathrm{Pb}$ collisions at $\sqrt{s_{NN}}=8.16\,\mathrm{TeV}$ within $1.5<y<4.0$~\cite{LHCb:2023chicFeeddown} and (b) forward $pp$ collisions at $\sqrt{s}=7\,\mathrm{TeV}$ within $2.0<y<4.5$~\cite{LHCb:2012chicFeeddown}.}
    \label{fig:chicbr_ldme}
\end{figure*}

Turning to the $\chic{}$ feed down contribution to prompt-$\jpsi$ production, we first compute 
the fraction $F_{\chic{1,2}\to \jpsi}$ by using the data driven approach for the prompt $J/\psi$ cross 
sections, which are directly from the LHCb 
measurements~\cite{LHCb:2017promptJpsi,LHCb:2011promptJpsi}. We also retain the kinematic 
approximation, $p_{\perp}^{\jpsi}\simeq p_{\perp}^{\chic{}}$, between $\chi_c$ and $J/\psi$.
We find that only the predictions with the CGC set~1 (MV) can reasonably describe both the $pp$ 
and $pA$ collisions data, as illustrated in Fig.~\ref{fig:chicbr_ldme}. The 
remaining two CGC sets systematically underestimate the data and the deviations tend to 
increase toward larger $p_\perp$. To quantify these trends, we calculate the $\chi^2/\mathrm{d.o.f.}$ for all the three parameter sets that are summarized in Tab.\ref{tab:chi2}.

\begin{table}[!htbp]
    \caption{$\chi^2/\mathrm{d.o.f.}$ for the data driven $\chic{}$ feed down fractions using the three CGC inputs.  Boldface marks the minimum in each row.}
    \label{tab:chi2}
    \centering
    \footnotesize
    \renewcommand{\arraystretch}{1.12}
    \setlength{\tabcolsep}{4pt}
    \begin{tabular}{|c|c|c|c|}
        \hline
        Data comparison & CGC set~1 & CGC set~2 & CGC set~3 \\
        \hline
        $p\mathrm{Pb}$
            & $\boldsymbol{0.94}$ & $5.46$ & $13.44$ \\
        $pp$
            & $\boldsymbol{1.20}$ & $2.61$ & $12.66$ \\
        \hline
        Combined
            & $\boldsymbol{1.06}$ & $4.17$ & $13.08$ \\
        \hline
    \end{tabular}
\end{table}

\begin{figure*}[!tp]
    \centering
    \captionsetup[subfloat]{farskip=2pt}
    \subfloat[]{
        \includegraphics[width=0.48\textwidth]{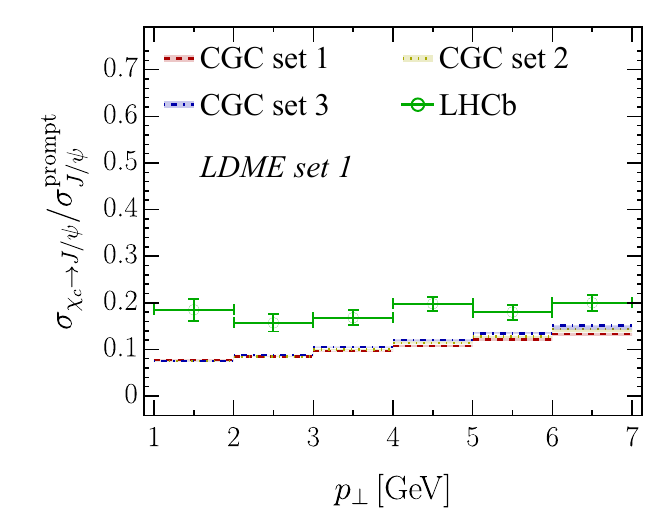}
        \label{fig:jpsi_pa_a}
    }
    \hfill
    \subfloat[]{
        \includegraphics[width=0.48\textwidth]{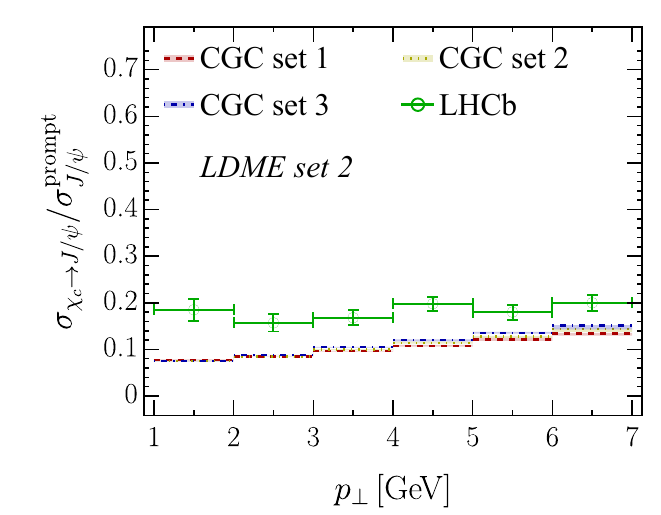}
        \label{fig:jpsi_pa_b}
    }
    \\
    \vspace{0.1cm}
    \subfloat[]{
        \includegraphics[width=0.48\textwidth]{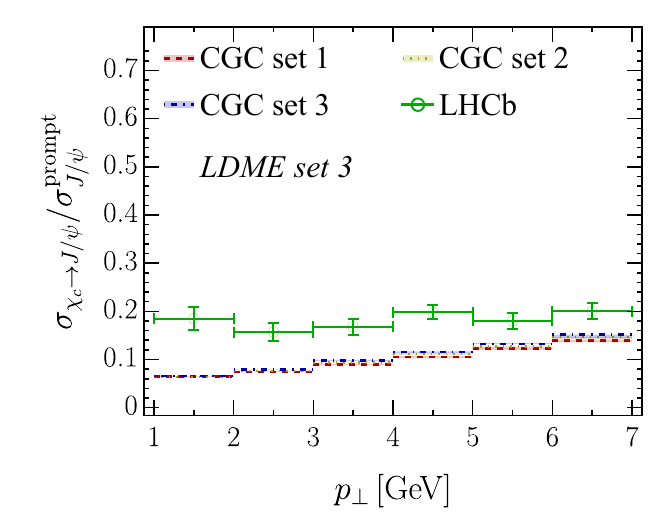}
        \label{fig:jpsi_pa_c}
    }
    \hfill
    \subfloat[]{
        \includegraphics[width=0.48\textwidth]{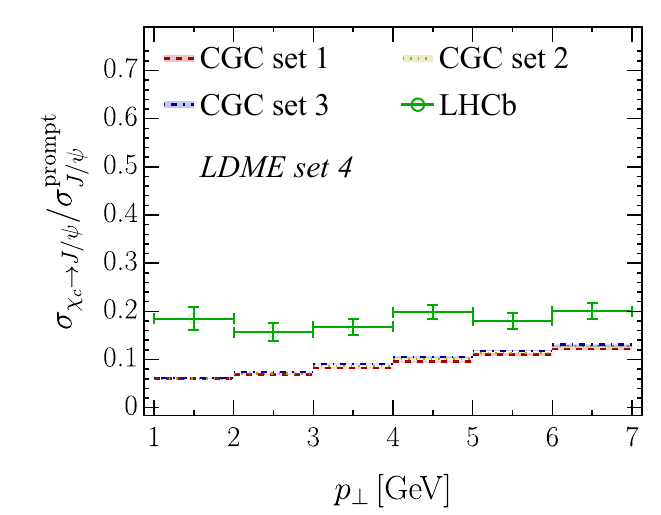}
        \label{fig:jpsi_pa_d}
    }
    \\
    \vspace{0.1cm}
    \subfloat[]{
        \includegraphics[width=0.48\textwidth]{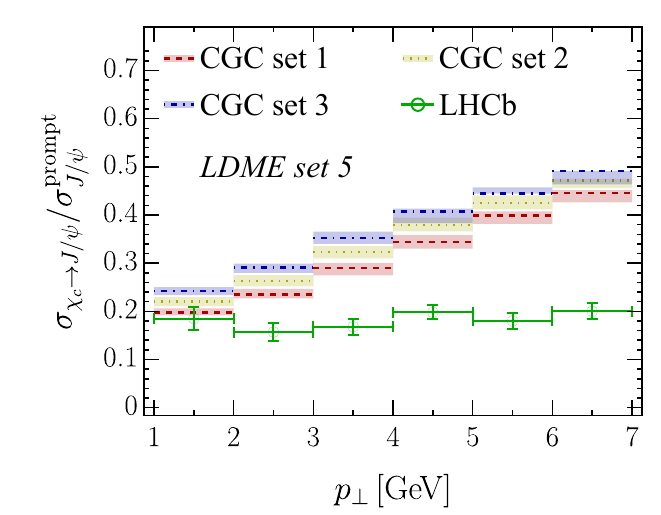}
        \label{fig:jpsi_pa_e}
    }
    \hfill
    \subfloat[]{
        \includegraphics[width=0.48\textwidth]{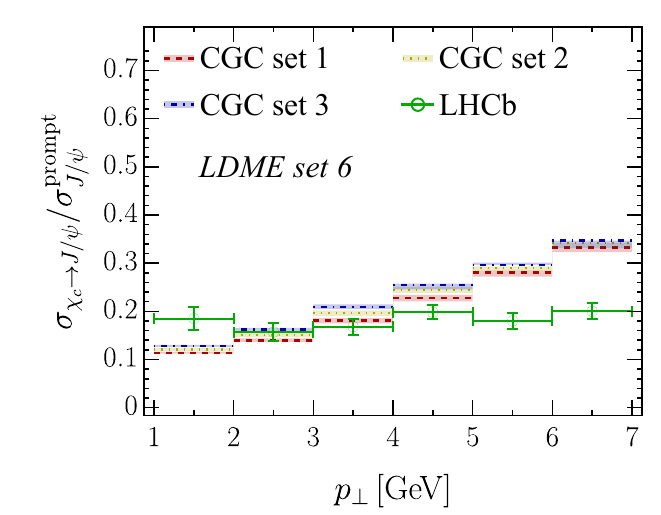}
        \label{fig:jpsi_pa_f}
    }
    \caption{Comparison of CGC+NRQCD predictions for the theoretical feed down fractions $F_{\chic{1,2}\to \jpsi}$ with the LHCb measurements in forward $p\mathrm{Pb}$ collisions at $\sqrt{s_{NN}}=8.16\,\mathrm{TeV}$ within $1.5<y<4.0$~\cite{LHCb:2023chicFeeddown}. Panels (a)--(f) use $\jpsi$ LDME sets~1--6 of Table~\ref{tab:ldme}, respectively, with all three CGC inputs shown in each panel.}
    \label{fig:jpsi_pa_ldme}
\end{figure*}

\begin{figure*}[!htbp]
    \centering
    \captionsetup[subfloat]{farskip=2pt}
    \subfloat[]{
        \includegraphics[width=0.48\textwidth]{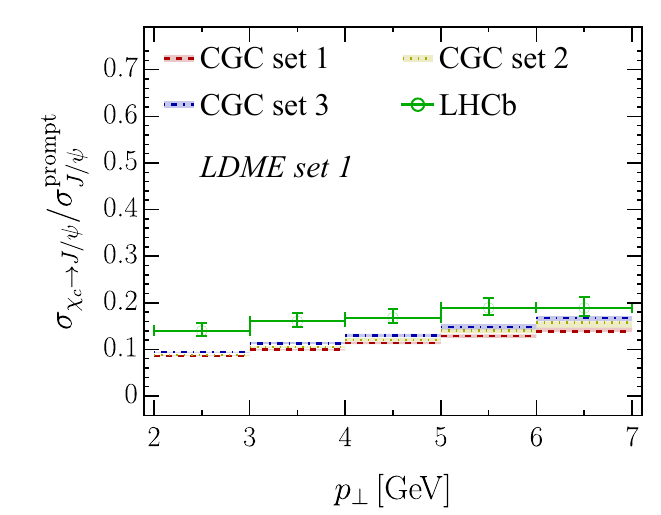}
        \label{fig:jpsi_pp_a}
    }
    \hfill
    \subfloat[]{
        \includegraphics[width=0.48\textwidth]{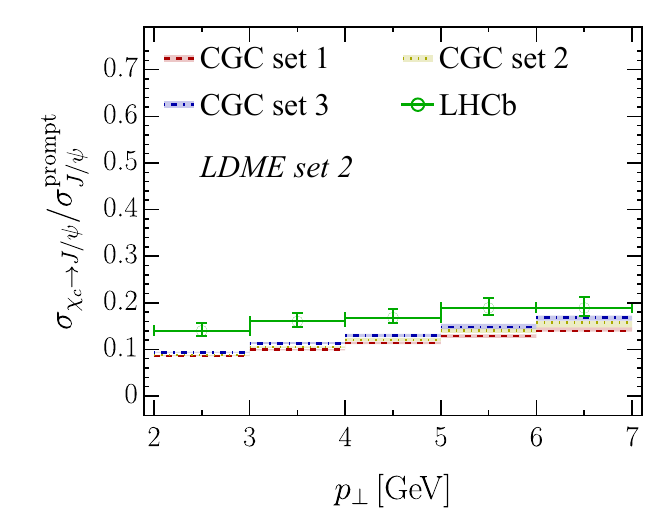}
        \label{fig:jpsi_pp_b}
    }
    \\
    \vspace{0.1cm}
    \subfloat[]{
        \includegraphics[width=0.48\textwidth]{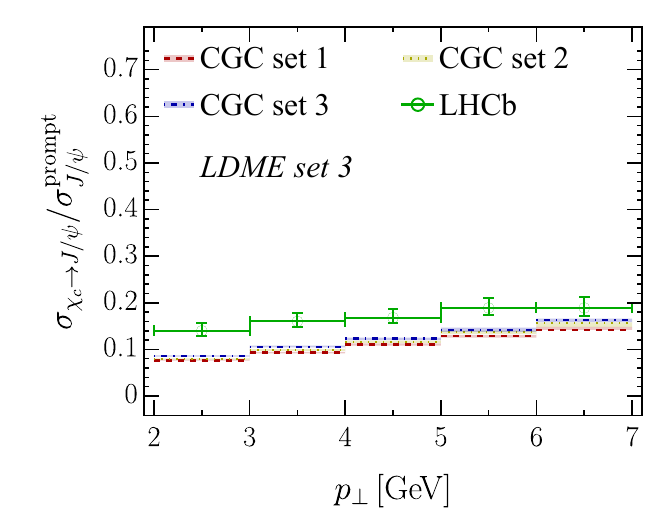}
        \label{fig:jpsi_pp_c}
    }
    \hfill
    \subfloat[]{
        \includegraphics[width=0.48\textwidth]{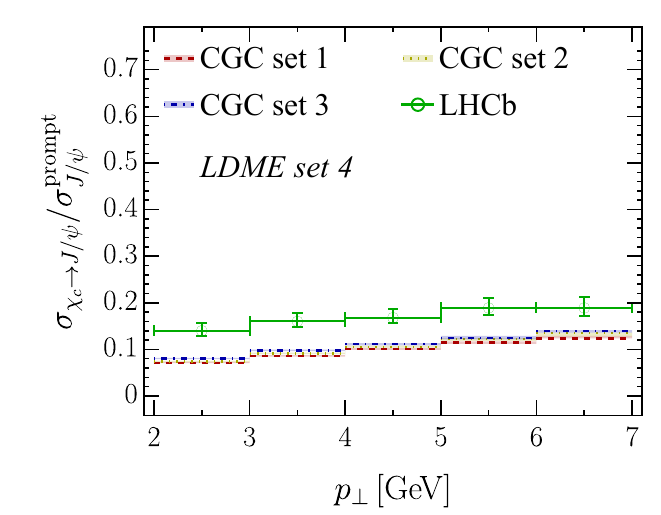}
        \label{fig:jpsi_pp_d}
    }
    \\
    \vspace{0.1cm}
    \subfloat[]{
        \includegraphics[width=0.48\textwidth]{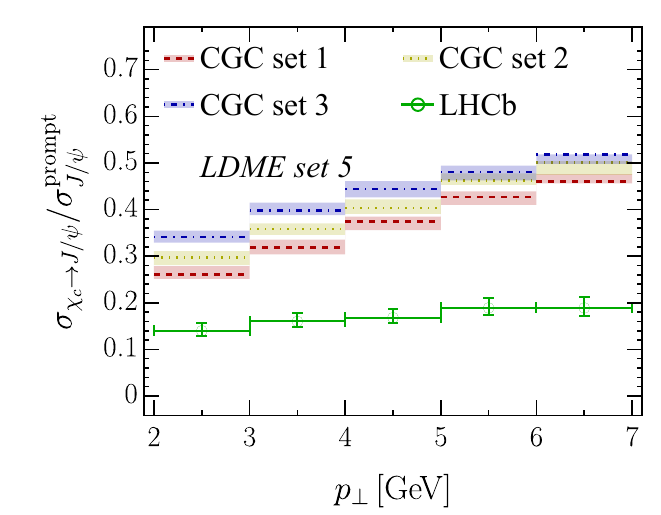}
        \label{fig:jpsi_pp_e}
    }
    \hfill
    \subfloat[]{
        \includegraphics[width=0.48\textwidth]{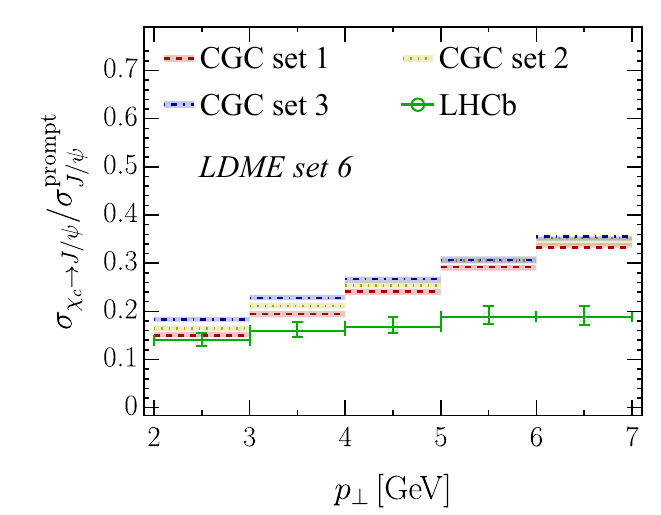}
        \label{fig:jpsi_pp_f}
    }
    \caption{Comparison of CGC+NRQCD predictions for the theoretical feed down fractions $F_{\chic{1,2}\to \jpsi}$ with the LHCb measurements in forward $pp$ collisions at $\sqrt{s}=7\,\mathrm{TeV}$ within $2.0<y<4.5$~\cite{LHCb:2012chicFeeddown}. Panels (a)--(f) use $\jpsi$ LDME sets~1--6 of Table~\ref{tab:ldme}, respectively, with all three CGC inputs shown in each panel.}
    \label{fig:jpsi_pp_ldme}
\end{figure*}

We further examine whether there exists a common choice of inputs that can describe both 
$\chic{}$ and prompt $\jpsi$ production within the CGC+NRQCD framework. For the direct 
$J/\psi$ production, we study with six $\jpsi$ LDME sets in Table~\ref{tab:ldme}.  Analogous to 
the behavior observed in $R_{\chi_{c}}$ case, the dipole and scale dependencies are partially 
canceled, and our predictions are rendered in Figs.~\ref{fig:jpsi_pa_ldme} and \ref{fig:jpsi_pp_ldme} to compare with the LHCb measurements in $p\mathrm{Pb}$ and $pp$ 
collisions, respectively. Panels (a)--(f) correspond to the predictions made by the six LDME 
sets each with all three CGC inputs shown. Unlike data driven approach, the good agreement with 
the data disappears when choosing the same MV input. Moreover, the pure theoretical predictions 
are also very sensitive to the values of the CO LDMEs: LDME sets~1--4 yield broadly similar 
central predictions that systematically underestimate the measured fractions in both collision 
systems, whereas LDME sets~5 and~6 result in fractions that increase with $p_\perp$ and 
overshoot the data at higher transverse momenta. Since the $\chic{}$ cross section in the 
numerator remains unchanged, this disagreement indicates a large discrepancy between the 
calculated and measured prompt-$\jpsi$ spectra.

Modifying the CGC inputs to improve the description of prompt $\jpsi$ production, however, will 
not resolve such tension. For instance, CGC set~2 together with LDME set~1 can 
describe the prompt $\jpsi$ spectrum in $pp$ collisions reasonably well over the
region $3<p_\perp<7\,\mathrm{GeV}$. Considering only the experimental uncertainties, the 
$\chi^2/\mathrm{d.o.f.}$ for these four bins is $1.73$ for CGC set~2, compared to $34.91$ for 
the MV initial condition. Conversely, CGC set~2 underestimates the $\chic{}$ contribution at 
larger $p_\perp$, as discussed above. The plots for CGC sets~2 and~3 in 
Figs.~\ref{fig:jpsi_pa_ldme} and \ref{fig:jpsi_pp_ldme}, also imply that changing the initial 
condition has only a modest effect on the theoretical feed down fractions and is unable to 
restore agreement over the full measured $p_\perp$ range. Consequently, none of the six LDME sets, combined with any of the three CGC inputs, reproduces the measured feed down fractions over the entire $p_\perp$ range within the theoretical and experimental uncertainties. Therefore, at QCD and $v^2$ LO, it seems not possible to describe $\chic{}$ and prompt $\jpsi$ production simultaneously. Since the NLO QCD and relativistic corrections are known to be crucial in collinear factorization calculations, the persistent tension between the CGC+NRQCD predictions and LHCb measurements may be resolved by incorporating these corrections in the future.

\section{Summary}\label{sec:conclusions}

To summarize, in this work, we have presented a systematic study of forward $\chic{J}$ 
production in $pp$ and $pA$ collisions within the CGC+NRQCD framework at LO in both $\as$ and 
$v^2$. Guided by a comparison of the dipole based ansatz with large $N_c$ and finite $N_c$ 
Gaussian treatments of the CS multipoint correlators, we adopt the finite $N_c$ Gaussian 
prescription as our default approach, which provides a common construction of these correlators 
while retaining their exact color structure. Under this prescription, we find that the CS 
channel contributes substantially to the total yield, accounting for around $45$--$90\%$ of 
$\chic2$ production and more than $60\%$ of $\chic1$ production over the investigated $p_\perp$ 
range. These sizable CS fractions differ substantially from the approximately $15\%$ CS 
contribution reported for $p_\perp$-integrated $\jpsi$ production, highlighting $\chic{1,2}$ 
production as a highly sensitive probe of the multipoint color correlations of the target.

When confronted with forward LHCb measurements, our predictions for the cross section ratio 
$R_{\chi_c}$ align well with the $p\mathrm{Pb}$ data within experimental uncertainties, but
systematically overshoot the $pp$ results across the measured $p_\perp$ range. We further 
examined the feed down fractions $F_{\chic{1,2}\to \jpsi}$ by using both data driven and purely 
theoretical schemes. While the data driven approach broadly reproduces the measurements
with the MV input, such agreement is lost when the denominator of prompt $\jpsi$ production is 
calculated within the same framework, implying a mismatch between the predicted and measured 
prompt $\jpsi$ spectra. On the other hand, improving the prompt $\jpsi$ description with an 
alternative CGC input will result in predictions for $\chic{}$ production that are inconsistent 
with LHCb data. Taken together, these comparisons demonstrate that none of the tested 
combinations of CGC inputs and $\jpsi$ LDME sets can provide a satisfactory simultaneous 
description of $\chic{}$ and prompt $\jpsi$ production, highlighting a persistent challenge in 
achieving a unified description across different charmonium states within the current framework 
at LO. Combined with the overestimation of $R_{\chi_c}$ in $pp$ collisions, these results 
motivate us to include higher order QCD and relativistic corrections within CGC+NRQCD in 
further study.

\allowdisplaybreaks

\section*{Acknowledgments} 
We thank Yan-Qing Ma, Jian Zhou, and Zhong-Bo Kang for helpful communications and discussions. 
This work is supported in part by the National Natural Science Foundation of China through 
Grant No.~12305082 and by the Fundamental Research Funds for the Central Universities
through Grant No. buctrc202432.

\appendix

\begin{widetext}
\section{\texorpdfstring{Form of $\tilde{\mathcal{H}}^\kappa_1$ in the dipole based ansatz}{Form of the momentum-space kernel in the dipole based ansatz}}
\label{app:quadrupole-approximation}
In this appendix, we present the general form of the momentum-space kernel $\tilde{\mathcal{H}}^\kappa_1$ entering Eq.~\eqref{eq:dsktCS1}:

\begin{multline}\label{eq:Hsinglet}
\tilde{\mathcal{H}}^\kappa_1 = (2\pi)^2 \mathcal{N}(\vt{k})\mathcal{N}(\vt{p}-\vt{k}) \Big[\delta^{(2)}(\vt{k}-\vtp{k}) - \delta^{(2)}(\vt{k}+\vtp{k}-\vt{p})\Big] W^\kappa(\vt{k},\vtp{k}) \\*
+ \frac{1}{2} \mathcal{N}(\vt{k})\mathcal{N}(\vtp{k})\mathcal{N}(\vt{p}-\vt{k}-\vtp{k}) \Big[W^\kappa(\vt{k},\vt{p}-\vt{k}) + W^\kappa(\vt{p}-\vtp{k},\vtp{k}) - W^\kappa(\vt{k},\vtp{k}) - W^\kappa(\vt{p}-\vt{k},\vt{p}-\vtp{k})\Big] \\*
- \frac{1}{2} \mathcal{N}(\vt{k})\mathcal{N}(\vtp{k})\mathcal{N}(\vt{p}) \Big[W^\kappa(\vt{k},\vt{p}-\vtp{k}) + W^\kappa(\vt{p}-\vt{k},\vtp{k}) - W^\kappa(\vt{k},\vtp{k}) - W^\kappa(\vt{p}-\vt{k},\vt{p}-\vtp{k})\Big] \;.
\end{multline}

The reflection symmetry of $W^\kappa(\vt{k},\vtp{k})$ follows from the pair's $C$ parity. Under $\vt{k}\to\vt{p}-\vt{k}$, it reads
\begin{equation}
W^\kappa(\vt{k},\vtp{k}) = (-1)^{L+S+1} W^\kappa(\vt{p}-\vt{k},\vtp{k}) \;.
\end{equation}
An analogous relation holds under $\vtp{k}\to\vt{p}-\vtp{k}$. For the $\CS3S1$ state, the hard factor is even under either reflection, so the first and third terms in Eq.~\eqref{eq:Hsinglet} make no contribution to the cross section. Equation~\eqref{eq:dsktCS1} then reduces to
\begin{equation}\label{eq:dsktCS}
\frac{\dd \hat{\sigma}_{\mathrm{CS}}^{\CS3S1}}{\dd^2\vt{p} \dd y} = \frac{\as S_\perp}{4(2\pi)^{5}(N_c^2-1)} \int \dd^2\vt{k} \, \dd^2\vtp{k} \, x_{p} f_{g/p}\left(x_{p}, \mu^{2}\right) \mathcal{N}(\vt{k})\mathcal{N}(\vtp{k})\mathcal{N}(\vt{p}-\vt{k}-\vtp{k}) \, \tilde{\mathcal{G}}^{\CS3S1}_1 \;,
\end{equation}
with
\begin{equation}
\tilde{\mathcal{G}}^{\CS3S1}_1 = \frac{1}{2}\Big[W^{\CS3S1}(\vt{k},\vt{k}) + W^{\CS3S1}(\vtp{k},\vtp{k}) - 2 W^{\CS3S1}(\vt{k},\vtp{k})\Big] \;.
\end{equation}
The cross section in Eq.~\eqref{eq:dsktCS} agrees with the proton-collinear limit of Eq.~(6) of 
Ref.~\cite{Ma:2014mri}. For intermediate states with charge conjugation parity $C=+$, including 
the $\CS3PJ$ channels of primary interest here, the cross section is evaluated from 
Eq.~\eqref{eq:dsktCS1} using the general kernel $\tilde{\mathcal{H}}_1^\kappa$ in 
Eq.~\eqref{eq:Hsinglet}.

\section{Coordinate-space CS hard factors}
\label{app:Gamma-singlet}

In this appendix, we collect the coordinate space hard factors $\Gamma_1^\kappa$ for the CS channels considered here. We use $r_\perp=|\vt{r}|$ and $r'_\perp=|\vtp{r}|$, and $K_n$ denotes the modified Bessel function of the second kind of order $n$.

For the $S$-wave $\CS3S1$ channel, the momentum-space hard factor takes the form
\begin{equation}
W^{\CS3S1}(\vt{l},\vtp{l}) = \frac{2m}{3 \tilde{X}_{l_\perp} \tilde{X}_{l'_\perp}} \;,
\end{equation}
where $\tilde{X}_{l_\perp} = l_{\perp}^2+m^2$ and $\tilde{X}_{l'_\perp} = {l'}_{\perp}^2+m^2$. The two-dimensional Fourier identity
\begin{equation}
\frac{1}{2\pi} \int \dd^2\vt{l} \, \frac{ e^{i \vt{l}\cdot\vt{r}}}{\tilde{X}_{l_\perp}} = K_0\left(m r_\perp \right)
\end{equation}
then yields the coordinate-space hard factor defined in Eq.~\eqref{eq:Gamma1_def}:
\begin{equation}\label{eq:Gamma-3S11}
\Gamma_{1}^{\CS3S1} = \frac{2m}{3} K_0\left(m r_\perp \right) K_0\left(m r'_\perp \right) \;.
\end{equation}

For the $P$-wave channels, the same procedure yields

\begin{align}
\Gamma_1^{\CS3P0}
={}&\frac{\vt{r}\!\cdot\!\vtp{r}}{3m r_\perp r'_\perp}
 \Big[mr_\perp K_0(mr_\perp)+K_1(mr_\perp)\Big]
 \Big[mr'_\perp K_0(mr'_\perp)+K_1(mr'_\perp)\Big] \;.
\label{eq:Gamma-3P01}
\\[0.5em]
\Gamma_1^{\CS3P1}
={}&\frac{1}{12m}\Bigg\{
 K_0(mr_\perp)K_0(mr'_\perp)
 \big(\vt{p}\!\cdot\!\vt{r}\big)
 \big(\vt{p}\!\cdot\!\vtp{r}\big)
 \notag\\*
&\quad
 +2\frac{\vt{r}\!\cdot\!\vtp{r}}{r_\perp r'_\perp}
 \Big[mr_\perp K_0(mr_\perp)-2K_1(mr_\perp)\Big]
 \Big[mr'_\perp K_0(mr'_\perp)-2K_1(mr'_\perp)\Big]
 \Bigg\} \;.
\label{eq:Gamma-3P11}
\\[0.5em]
\Gamma_1^{\CS3P2}
={}&\frac{1}{60m}\Bigg\{
 3K_0(mr_\perp)K_0(mr'_\perp)
 \Big[\big(\vt{p}\!\cdot\!\vt{r}\big)
 \big(\vt{p}\!\cdot\!\vtp{r}\big)
 +4m^2\big(\vt{r}\!\cdot\!\vtp{r}\big)\Big]
 \notag\\*
&\quad
 +2\frac{\vt{r}\!\cdot\!\vtp{r}}{r_\perp r'_\perp}
 \Big[mr_\perp K_0(mr_\perp)-2K_1(mr_\perp)\Big]
 \Big[mr'_\perp K_0(mr'_\perp)-2K_1(mr'_\perp)\Big]
 \Bigg\} \;.
\label{eq:Gamma-3P21}
\end{align}

\end{widetext}

\bibliographystyle{apsrev4-1}
\bibliography{refs}

\end{document}